# Deterministic integrated tuning of multi-cavity resonances and phase for slow-light in coupled photonic crystal cavities

Tingyi Gu[1*], Serdar Kocaman[1], Xiaodong Yang[2], James F. McMillan[1], Mingbin Yu[3], Guo-Qiang Lo[3], Dim-Lee Kwong[3], and Chee Wei Wong[1*]

[1]*Optical Nanostructures Laboratory, Center for Integrated Science and Engineering, Solid-State Science and Engineering, and Mechanical Engineering, Columbia University, New York, NY* 10027

[2]*University of California at Berkeley and Lawrence Berkeley National Laboratory, Berkeley, CA* 94720

[3]*The Institute of Microelectronics, 11 Science Park Road, Singapore Science Park II Singapore* 117685*, Singapore*

We present the integrated chip-scale tuning of multiple photonic crystal cavities. The optimized implementation allows effective and precise tuning of multiple cavity resonances (up to ~1.60 nm/mW) and inter-cavity phase (~ 0.038 $\pi$/mW) by direct local temperature tuning on silicon nanomembranes. Through designing the serpentine metal electrodes and careful electron-beam alignment to avoid cavity mode overlap, the coupled photonic crystal *L3* cavities preserve their high quality factors. The deterministic resonance and phase control enables switching between the all-optical analogue of electromagnetically-induced-transparency (EIT) to flat-top filter lineshapes, with future applications of trapping photons/photonic transistors and optoelectronic modulators.

Based on analogies between classical electromagnetic fields and quantum probability amplitudes in atomic physics, EIT and its photonic correspondence have been examined in atomic three-level canonical systems [1-2], atom – optical cavity systems [3], and chip-scale coupled photonic resonators such as indirectly coupled whispering gallery resonators [4-5] and photonic crystal cavities [6-7]. In its optical analogue, the interferences of the electromagnetic wave between excitation pathways to the upper level in three-level realization has led to highly-dispersive absorption cancellation of the medium [3], resulting in phenomena such as stopping and dynamical storage of light [8-9].

Here we present the observations of deterministic resonance and phase tuning of multiple photonic crystal cavities with precisely-positioned chip-scale integrated electrodes, followed by the realization of an all-optical solid-state analogue to EIT on-chip. The optical EIT-like lineshape has a comparable bandwidth-delay product to the atomic systems, although it has a significantly larger bandwidth and a correspondingly shorter delay [8]. To achieve the coherent interferences on-chip, the detuning and phase mismatch between the optical transitions or oscillators must be tightly controlled; to overcome the resonance variations between multiple cavities on-chip [10], tuning schemes involving for example optical [11] or electrically carrier injection [12], atomic layer deposition [13], integrated piezoelectric [14] and thermal heating [15] have been examined. Efficient carrier injection through Drude plasma dispersion has enabled for example fast CMOS-compatible optoelectronic modulators with appreciable extinction ratios and small drive voltage requirements [16]. These integrated approaches pave a scalable approach for chip-scale tuning, such as to simplify the resonance alignments of multiple cavities [17] in variable delay lines, controllable light-matter interactions in slow light photonic crystal waveguides [18], and high-speed efficient optical interconnects and transceivers on-chip [16].

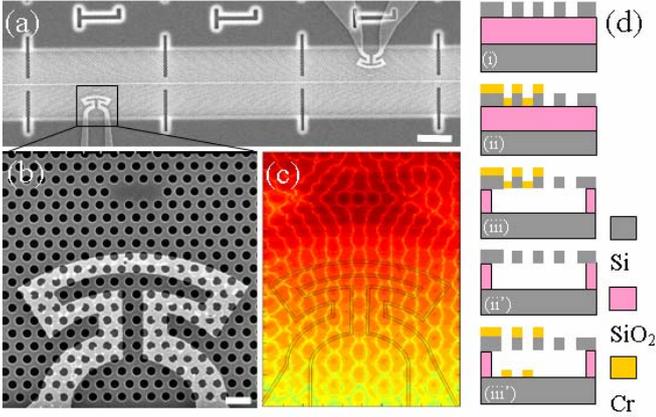

FIG. 1. Chip-scale integrated tuning of photonic crystal two-cavity optical EIT system. (a) SEM of thermally-tuned coupled cavities with thermal isolation trench and tuning electrodes. Scale bar: 5 μm. (b) SEM of single cavity. Scale bar: 500 nm. (c) 3D FDTD simulated model profile (log scale) with outline of thermal tuning electrode. (d) Schematics of two-step nanofabrication process flow. (i) SOI wafer, (ii) electron-beam lithography defined tuning electrodes, (iii) suspended silicon membrane with electrodes. (ii') initial sacrificial release, following by (iii') electron-beam writing of the electrodes.

The photonic crystal cavities and membranes examined in this work are fabricated on a 250 nm thick silicon-on-insulator device layer via optimized 248 nm deep-ultraviolet lithography and etching for reduced disorder scattering [19]. The lattice constant of the photonic crystal is 420 nm with 110 nm hole radius ($a$ = 420 nm, $r$ = 0.261$a$, $t$ = 0.595$a$). Each cavity is designed with three missing central holes (termed "$L3$"), with the nearest neighboring holes tuned and shifted [20] to enhance intrinsic quality factor ($Q_{in}$ ~ 60,000 from 3D FDTD calculations [6]). To achieve the integrated tuning electrodes on the suspended membranes, two nanofabrication approaches are developed. The first approach is shown in Fig. 1d subpanels (i), (ii), (iii), where the electrode features are first aligned to the photonic crystal cavities through electron-beam lithography and the sacrificial undercut etch of the photonic crystal membranes as the final step. The second approach involves sacrificial release of the membranes first, before electron-beam alignment and electrode patterning through lift-off [Fig. 1d subpanels (i), (ii'), (iii')]. The misalignment in the spatial electron-beam registration between the electrodes and the cavities is estimated to be ~ 200 nm. The tuning electrode is electron-beam evaporated with 100 nm chrome. The folded serpentine layout of the heating electrodes, along with a convex profile "bending away" from the cavity center, is designed and written so as to maximize the (joule) heating while minimizing perturbation to the $L3$ optical cavity field (from intrinsic metal absorption) when positioned close to the nanocavities [15]. The symmetric electrode profile ensures symmetric heat flux to the coupled cavities. With a brief HF etch before electron-beam evaporation, the native oxide or organic materials on the top surface is removed and hence direct contact between the chrome and photonic crystal membrane is achieved. With the second approach in nanofabrication [Fig. 1d subpanels (i), (ii'), and (iii')], the resulting edge interface of the lifted-off electrodes is around 50 nm and the resistance for this specific design is around 1.3 kΩ.

Transmission measurements are performed with amplified spontaneous emission sources, with polarization controllers and tapered lensed input/output fiber coupling. DC bias is applied to the nanofabricated electrodes. The output is sent to an optical spectrum analyzer. In these measurements, each cavity is implemented in the over-coupled regime ($Q_{coupling}$ ~ 6,000), with a resulting measured loaded $Q$ in the range of 5,600 to 8,500 to allow for coherent in-plane cavity-cavity interference. The typical resonance extinction ratio is measured to be ~ 15 dB, and the correspondent intrinsic cavity $Q$s is ~ 33,000. Fig. 1c shows an example 2D finite-difference time-domain simulation of the $L3$ cavity with the tuning electrode outline. We note that the thermal electrodes are nanofabricated at four lattice periods away from the cavity (Fig. 1c) such that the heat is effectively diffused from electrodes to the cavity region and intrinsic cavity $Q$ is not affected by the metal absorption. If the metal is placed too close to cavity (less than three lattice periods), much of the light would be coupled into plasmon modes (experimentally observed from the top scattering image) of metal electrode and the cavity $Q$ is significantly attenuated. The distance separation ($L$) between two $L3$ cavities is 60 μm and includes thermal isolation trenches to achieve independent cavity and phase (in the photonic crystal waveguide) tuning.

Fig. 2a shows the transmission spectra of two $L3$ cavities when the shorter wavelength resonance (at 1581.9 nm) of the active cavity is thermally red-shifted to align up to the longer wavelength resonance (1584.4 nm) of the target one.

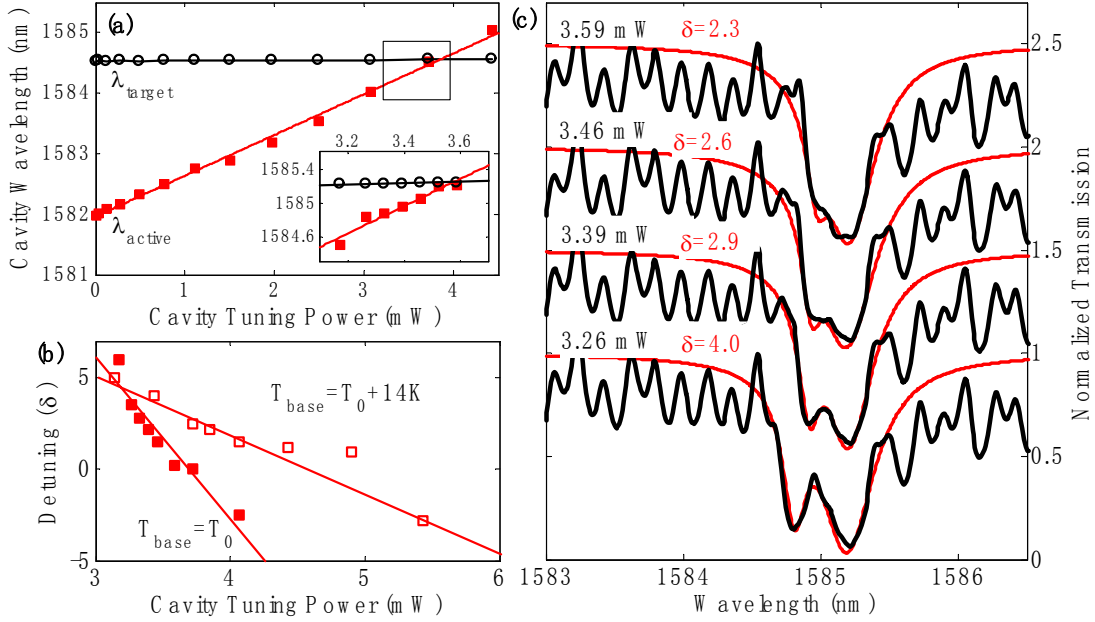

FIG. 2. Integrated tuning of the two-cavity resonances. (a) The resonant wavelength of the active cavity (solid squares) and the target cavity (empty circles) versus the tuning power. Inset: fine tuning cavity resonances in the cavity-cavity interference region. (b) Detuning versus local tuning power within the same region as inset of (a). Solid and empty squares are data measured at base temperature (T0) and T0+14K respectively. Lines are linear fits. (c) Example two-cavity coherent transmission under different local tuning powers (δ = 2.3, 2.6, 2.9 and 4.0, correspondent to cavity tuning power at 3.59 mW, 3.46 mW, 3.39 mW and 3.26 mW respectively). Solid (black) lines are experimental data and the dashed (red) lines are coupled-mode theory fits (Offset: -0.5).

In Fig. 2a inset, we show the fine-tuning at a higher base temperature (with ~ 1 nm red-shifted resonances). In Fig. 2b we plot the fine tuning of the active cavity near the target cavity for observing the interference patterns, with the resulting detuning $\delta$ [$=2\tau_{total}(\omega_1-\omega_2)$] illustrated. We emphasize that when $\delta < \sim 3.5$, the linewidth of the transparency peak is narrower than the individual cavity linewidths [6, 21], in the regime of Fano- and EIT-like interferences. We also exploring the sensitivity of interference lineshapes to different chip (base) temperatures – the tuning efficiency $d\delta/dP$ drops from 8.5 /mW to 2.9 /mW after the base temperature being raised by 14K. Fig. 2c shows four example transmission lineshapes with interference between two indirectly coupled cavities, with detuning $\delta$ of 2.3, 2.6, 2.9 and 4.0, corresponding to local cavity tuning power 3.59, 3.49, 3.39 and 3.26 mW applied on the electrodes respectively. In addition to the lineshape shown in Fig. 2c with $\delta$ at 4.0, the other panels also illustrate an even smaller detuning through the integrated control, where an asymmetric lineshape indicative of Fano-like interferences is observed. There is also a Fabry-Perot background in the measured spectra which arises from the finite termination of the photonic crystal and the chip. By simulating the experimental data with coupled mode theory (CMT), the cavity-to-cavity phase difference that satisfies the condition of forming a Fabry–Perot resonance is $0.85\pi$ for all cases.

The thermal control of the cavity has a resonance redshift of 1.60 nm/mW at room temperature. When only the shorter wavelength cavity is thermally-tuned, we note that there is negligible cross-talk for the target cavity, which is observed to be 0.038 nm/mW by linear curve fitting the unintended target cavity shift (Inset of Fig. 2(a)). The devices typically operate with voltage biases up to ~ 6 V. The thermal impedance of the photonic crystal cavity is defined as: $1/R_{th}=\Delta T/\Delta P =(\Delta\lambda/\Delta P)\times(\Delta n/\Delta\lambda)\times(\Delta T/\Delta n)$, where $R_{th}$ is the thermal resistance (in units of mW/K), $\Delta T$ is the temperature difference, $\Delta P$ is the joule heating power supplied by electrodes, and $\Delta\lambda$ is the cavity wavelength shift. $\Delta\lambda/\Delta P$ is given by the measurement shown in Fig. 2a. The temperature dependence of the refractive index $\Delta n/\Delta T$ in silicon is $1.86\times10^{-4}$ /K [22], with $\Delta n/\Delta\lambda=n_{Si}/\lambda_{active}$ obtained within the first-order perturbation for the high index material [12, 23] ($n_{Si}$=3.485). The thermal resistance at room temperature is estimated as 18.7 K/mW, comparable to optical tuning at 15.4 K/mW [6, 21].

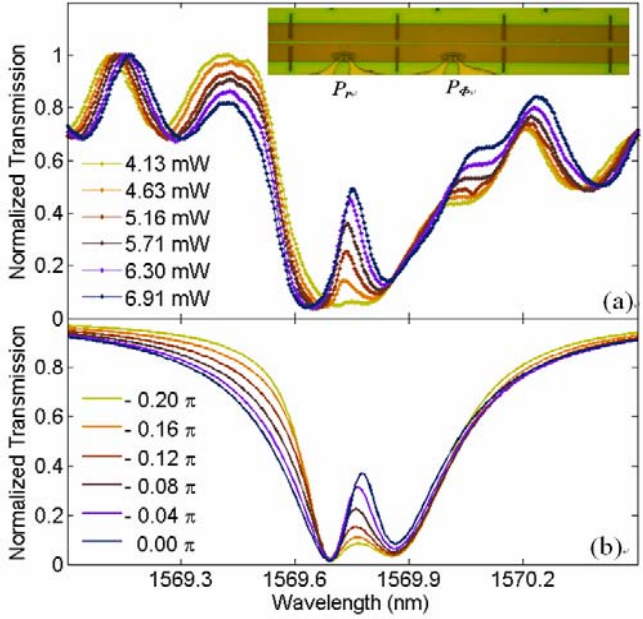

Fig.3. Transmission spectrum (a) Measurement with fixed resonance tuning power ($P_r$=40.8 mW) and increasing phase tuning power ($P_\Phi$). (b) Simulation with fixed cavity resonances and decreasing phase. Inset: optical image of nanofabricated resonance- and phase-tuning electrodes on photonic crystal membrane with thermal isolation trenches.

We next examine the controlled cavity-cavity phase tuning with integrated electrodes as illustrated in Fig. 3. Phase between the cavities is given by $\Phi = \int_0^L \omega_g n_{eff}(l,T)/c \, dl$ where $n_{eff}$ is local effective index of photonic crystal waveguide, 2.768 at 1550 nm for room temperature, and $L$ is the length of the photonic crystal waveguide between two cavities. A second electrode is placed beside the photonic crystal waveguide to locally adjust the refractive index of the waveguide between the two cavities (Fig. 3 inset). For this second sample, the active cavity has $Q_{loaded}$ = 7,500, $Q_{in}$ = 30,000, and resonance wavelength is thermally tuned to 1569.86 nm, and the target cavity has loaded quality factor $Q_{loaded}$ = 8,500, $Q_{in}$ = 15,000, with resonance wavelength at 1569.97 nm. After placing the resonant frequencies of the two cavities close enough ($\delta$ = 1.9 and less than 3.5 [6]), the relative cavity-cavity phase difference is adjusted through thermal-optic control of the waveguide between two cavities: $\Delta\Phi = dn/dT \times \Delta T \times L_{eff}/\lambda_c \times 2\pi$, where $\Delta T$ is the increased waveguide temperature, and $L_{eff}$ is the the length of the waveguide effectively heated up by the electrode placed between two cavities (right electrode in insert of Fig. 3 (a)).

Fig. 3a shows that the transmission lineshape is gradually tuned from a flat-top reflection filter to an EIT-like optical resonance when the power applied on integrated phase tuning electrode is increased from 4.13 mW to 6.91 mW. The phase shift $\Delta\Phi$ by the local refractive index change is tuned from $0.07\pi$ to $-0.03\pi$ when the phase local tuning power increases from 4.13 mW to 6.91 mW, which corresponds to a sensitivity of 0.038 $\pi$/mW. To illustrate the phase tuning physics, the CMT simulated lineshape with fixed detuning ($\delta$ at 1.6) and increasing phase (from out-of-phase to in-phase) is illustrated in Fig. 2b, and matched well with experimental results without any fitting parameters. With the average of the two cavity resonances maintained, the tilted EIT-like peak with increasing phase tuning power is induced by the different $Q$s of the two interfering cavities.

In summary we have demonstrated the integrated resonance tuning of multiple photonic crystal cavities by precisely electron-beam-positioned electrodes. The differential local cavity resonance tuning of 1.60 nm/mW and phase tuning of 0.038 $\pi$/mW have enabled coherent interferences from flat-top reflection to narrow band pass filters. This observed chip-scale control has applications in tunable delay lines, efficient modulators, and photon pulse trapping and release in scalable multi-cavity implementations.

The authors acknowledge discussions with J. Zheng, C. Husko, and L. Fetter, loaning of facilities from Prof. Shepard and Prof. Attinger, and Columbia cleanroom facilities. The project is supported by a NSF CAREER Award (ECCS-0747787) and partially supported by the Nanoscale Science and Engineering Initiative of the National Science Foundation under NSF Award Number CHE-0641523, and by the New York State Foundation for Science, Technology, and Innovation (NYSTAR).